\documentclass[preprint,12pt]{elsarticle}

\usepackage{lineno,hyperref}
\usepackage{natbib}
\usepackage{geometry}
\usepackage{fleqn}
\usepackage{graphicx}
\usepackage{newtxtext,newtxmath}
\usepackage{hyperref}
\usepackage{booktabs}
\usepackage{siunitx}
\usepackage{bm}
\usepackage{amsmath}
\usepackage{enumitem}
\usepackage{esvect}
\usepackage{subcaption}
\usepackage{soul,xcolor}
\modulolinenumbers[5]

\DeclareMathOperator*{\argmax}{argmax}

\graphicspath{{./figures/}}
\journal{Elsevier}

\begin{document}

\begin{frontmatter}

\title{\large{Data-Driven Prediction and Uncertainty Quantification of PWR Crud-Induced Power Shift Using Convolutional Neural Networks}}

\author[NCSU]{Aidan Furlong\corref{mycorrespondingauthor}}
\cortext[mycorrespondingauthor]{Corresponding author}
\ead{ajfurlon@ncsu.edu}

\author[NCSU]{Farah Alsafadi}

\author[NCSU]{Scott Palmtag}

\author[VCN]{Andrew Godfrey}


\author[NCSU]{Xu Wu}

\address[NCSU]{Department of Nuclear Engineering, North Carolina State University    \\ 
	Burlington Engineering Laboratories, 2500 Stinson Drive, Raleigh, NC 27695 \\}

\address[VCN]{Veracity Nuclear, Lenoir City, TN, USA}


\begin{abstract}
The development of Crud-Induced Power Shift (CIPS) is an operational challenge in Pressurized Water Reactors that is due to the development of crud on the fuel rod cladding. The available predictive tools developed previously, usually based on fundamental physics, are computationally expensive and have shown differing degrees of accuracy. This work proposes a completely ``top-down'' approach to predict CIPS instances on an assembly level with reactor-specific calibration built-in. Built using artificial neural networks, this work uses a three-dimensional convolutional approach to leverage the image-like layout of the input data. As a classifier, the convolutional neural network model predicts whether a given assembly will experience CIPS as well as the time of occurrence during a given cycle. This surrogate model is both trained and tested using a combination of calculated core model parameters and measured plant data from Unit 1 of the Catawba Nuclear Station. After the evaluation of its performance using various metrics, Monte Carlo dropout is employed for extensive uncertainty quantification of the model predictions. The results indicate that this methodology could be a viable approach in predicting CIPS with an assembly-level resolution across both clean and afflicted cycles, while using limited computational resources.
\end{abstract}


\end{frontmatter}


\section{Introduction}
 

The phenomenon of Crud-Induced Power Shift (CIPS) is an operational challenge in Pressurized Water Reactors (PWRs) due to the accumulation of borated crud, a combination of corrosion products and boron species. Also known as Axial Offset Anomaly (AOA), it presents as a deviation in the measured axial power shape from the calculated core model used to design the loading pattern and perform safety analyses \cite{deshon2011pressurized}. Since the axial offset of operating PWRs is a safety-related quantity, it is of interest to both understand and predict the behavior of this effect. Another key concern is the development of Crud-Induced Localized Corrosion (CILC), where the accumulation of crud reduces the ability of fuel rods to shed heat. This directly leads to an increase in the cladding temperature, increasing thermal stresses and accelerating the corrosion caused by chemical interactions. Crud-related fuel failures can result from CILC and CIPS, as reported by power plants such as the Seabrook Nuclear Power Plant and Palo Verde Nuclear Generating Station \cite{turinsky2016modeling}.


The development of borated crud requires three factors: subcooled nucleate boiling (SNB), corrosion products circulating the primary loop (such as iron and nickel species), and the presence of soluble boron in the coolant \cite{deshon2011pressurized}. The first contributor is SNB, which occurs at imperfection sites along the fuel cladding's exterior. These nucleation sites, above the midplane of the core, are the location at which the liquid coolant transforms into voids which rise and then collapse. These cladding areas also provide a suitable region for the deposition of corrosion products to occur \cite{short2013multiphysics}. Throughout a cycle, the SNB locations are dynamic and shift with the peak pin powers. The deposition process builds a highly porous structure that can exceed $80$ \si{\micro\meter} in thickness. Boiling chimneys also form through this layer where voids can become momentarily trapped against the cladding wall, raising the local wall temperature \cite{deshon2011pressurized}. Once the corrosion products have deposited on the surface of the cladding and the crud structure has formed, the local concentration of boron species rises, increasing neutron absorption and reducing local power.


Measuring the build-up of crud under operational conditions is challenging. Crud's higher solubility in cold temperatures, compared to hot temperatures, makes it difficult to measure during refueling. When the plant is shut down, crud layers dissolve, leaving only a fraction that can be measured. As a result, indirect methods are used to assess the presence and severity of CIPS mid-cycle. This typically involves comparing axial power shapes from in-core detectors with the expected values calculated for the cycle. One specific quantity is known as the ``CIPS Index'', proposed by Andrew Godfrey, which computes and compares the deviations along four axial spans of a given assembly \cite{CIPSIndexreport}. It is of note that assemblies without explicit CIPS are still affected by the phenomenon, and can similarly see significant axial deviations.

Various models have been introduced to investigate crud behavior by incorporating coupled physics, such as neutronics, thermal-hydraulics, and coolant chemistry. These mechanistic models utilize a ``bottom-up'' approach to predict crud growth. Notable examples of modeling and simulation tools used to analyze crud and CIPS behavior include the Boron-induced Offset Anomaly (BOA) risk assessment tool \cite{epri2010boron}, developed by the Electric Power Research Institute, and the MAMBA code \cite{turner2016virtual}, created as part of the Consortium for Advanced Light Water Reactor Simulation (CASL) program. MAMBA is an integral component of the Virtual Environment for Reactor Analysis (VERA) modeling suite. In the United States, BOA and MAMBA are employed to assess the potential for crud deposition and evaluate the risk of CIPS and CILC. These models, however, are computationally expensive to run and do not involve measurement-related data from specific reactors as inputs. 

As machine learning (ML) applications continue to expand, there is a recent focus on data-driven approaches aimed at creating fast-running reduced-order models with acceptable accuracy. In a prior study \cite{andersen2022novel}, efforts were concentrated on developing a computationally efficient model to integrate with a loading pattern optimization algorithm. The study introduced ``crUdNET'', a neural network surrogate model trained with MAMBA simulation data capable of predicting crud distributions using power and core-wide parameters. Another approach was proposed by Liao et al. to reduce the computational costs associated with BOA to support a high volume of crud mass predictions \cite{liao2022applying}. This study implemented a simple five-layer deep neural network to predict crud mass values using over $111,000$ BOA runs and $16$ input features. This approach produced crud predictions within $\pm 0.25$ \si{lbm} of the BOA reference. With the two aforementioned works focusing on the use of computational data from the mechanistic codes as a training basis, there are currently no known attempts at predicting CIPS with data-driven surrogate models while using real measurement-derived data from a PWR.

In this work, we present the use of artificial neural networks (ANNs) trained with real measurement data as an approach to predict whether a given assembly will be affected by CIPS at a given step in time, building on earlier work \cite{furlong2023machine} \cite{furlong2024predicting}. The computer vision technique of three-dimensional (3D) convolution is also implemented to retain spatial relationships in the input features and to increase the overall efficiency. This method focuses on using a combination of measurement-derived data and core model data obtained from the Catawba Nuclear Station Unit 1 (CNS-1) over the span of three cycles. The predictions are then evaluated in multiple frames to investigate trends in performance and error. Extensive uncertainty quantification (UQ) for the ML model using Monte Carlo dropout (MCD) was then implemented to assess the confidence and reliability in the model's predictions. 

The organization of the subsequent sections in this paper is as follows: Section \ref{sec:data-and-model} introduces the problem and details the data employed. Section \ref{sec:methods} outlines the general approach, architecture of the neural network used, and UQ methodology. In Section \ref{sec:results}, we present the findings of our investigation, encompassing the CIPS predictions and the associated UQ. Discussion about the practical implementation and future work is then provided in Section \ref{sec:discussion}. Finally, our concluding remarks on this study can be found in Section \ref{sec:conclusion}.

\section{Training Data and ML Model} \label{sec:data-and-model}

\subsection{Training Data}

\begin{figure}[htb]
    \centering
    \includegraphics[width = 0.8\textwidth]{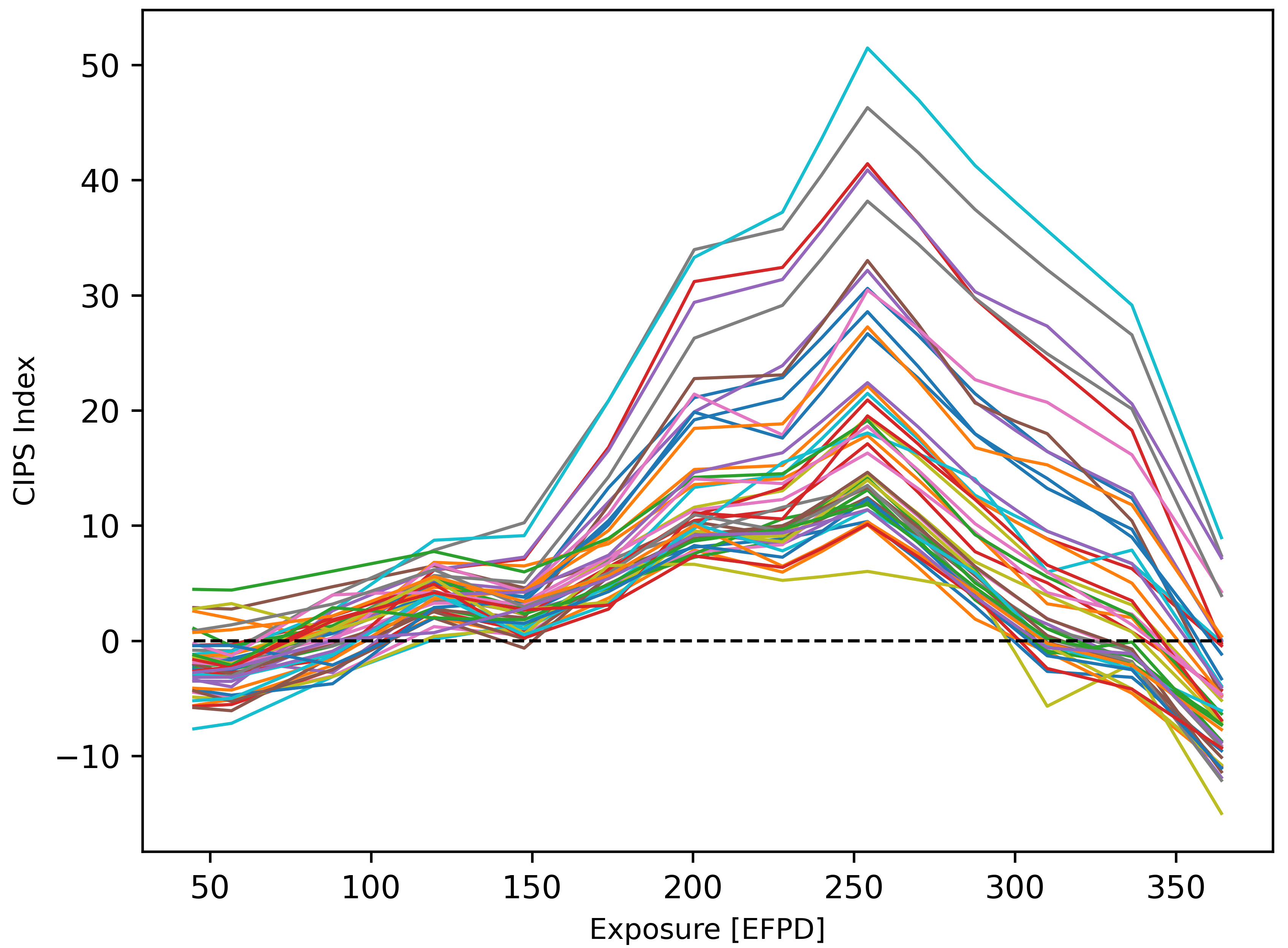}
    \caption{CIPS values for the 49 fuel assemblies over Cycle 8.}
    \label{fig:cycle8-dataset-cips}
\end{figure}

This work is based on data provided by Duke Energy for three cycles from the Catawba Nuclear Station Unit 1 (CNS-1): Cycles $6$, $7$, and $8$. Two types of files were given: \texttt{cX.h5} and \texttt{compX.h5}, where ``\texttt{X}'' means the cycle number (6-8). The former contained core models generated by the reactor analysis tool MPACT, which assumed quarter-core symmetry. The \texttt{compX.h5} files provided comparisons between in-core instrumentation and the relevant core model calculations. These files also contained the CIPS Index values, which are the focus of this study. Cycles $6$ and $7$ were ``clean'', and did not see any instances of CIPS. Cycle $8$ was significantly affected by CIPS, making it useful as the positive class' data. Due to the inherent data imbalance between the positive and negative classes, Cycle $6$ was excluded from the working dataset. The two cycles' that were included cover $49$ assembly positions at $24$ different time steps per cycle, leading to a combined total of $2352$ with some detectors inoperable at different times. Figure \ref{fig:cycle8-dataset-cips} illustrates the CIPS Index trajectories for each of the $49$ fuel assemblies throughout Cycle 8, with peak CIPS occurring at $254$ effective full power days (EFPDs).

The ML model is constructed with four parameters being considered. Three of these were MPACT-calculated parameters: the 3D pin power maps, coolant boron concentration, and core exposure in EFPD. The 3D pin powers were chosen due to their relation to SNB, and are unique to each assembly. The coolant boron concentration is included as a surrogate ``source term'', and the core exposure serves as a temporal component. Larger sets of input features were attempted, including parameters such as moderator temperature and core power, but the added complexity along with the limited nature of the dataset contributed to overfitting and undesirable performance. The output parameter is the probability that a given assembly will be affected by CIPS. Developing the training instances for this parameter involved thresholding the CIPS Index with a value of $10$ to segment positive and negative classes, since values above $10$ are indicative of ``possible CIPS'' \cite{CIPSIndexreport}. A listing of these features is included in Table \ref{tab:parametrization}.

\begin{table}[!htb]
	\centering
	\begin{tabular}{lcc}
		\hline
		\textbf{Input}  & \textbf{Units} & \textbf{Dimensions} \\ \hline
		Boron Concentration     & \si{ppm}  & [$2352$] \\
		Core Exposure           & \si{EFPD} & [$2352$] \\
        3D Pin Powers           & \si{-}    & [$2352,17,17,44$]  \\ \hline
        \textbf{Output}  & &  \\ \hline
        Probability of CIPS-positive & \si{-} & [$2352$] \\ \hline   
	\end{tabular}	
    \caption{Features and dimensions.}
	\label{tab:parametrization}
\end{table}

\subsection{ML Model for CIPS Prediction}
\label{sec:nn_background}

Considering the dataset and input features, the ML model was designed to include two distinct types of neural networks: multilayer perceptron (MLP), also known as deep neural network, and convolutional neural network (CNNs). These two networks differ in their architecture and suitability for handling various data types.

MLPs consist of an input layer, one or more hidden layers, and an output layer. In these fully-connected feed-forward neural networks, each neuron in a specific layer is connected to every neuron in the subsequent layer. The learning process of these networks incorporates the backpropagation method, wherein the network parameters (weights and biases) are updated. These parameters are optimized using the gradient descent method to minimize the difference between the true output and the model’s predictions. This type of network is suitable for handling various types of data, such as coolant boron concentration and exposure in our case. However, since MLPs take inputs as a vector and do not capture spatial information, they are not appropriate for handling 3D pin powers where neighboring relationships are likely crucial. To account for this, the CNN was considered.

CNNs have gained widespread adoption in computer vision approaches, with applications such as handwriting detection and object recognition. Originally based on the feline visual cortex, convolutional layers allow for pattern recognition using the geometric similarity between relative entry locations within data rather than absolute \cite{hubel1962receptive} \cite{fukushima1980neocognitron}. CNNs have also gained wide attention in the energy domain. For example, in nuclear energy, CNN-based fault analysis and diagnosis models have been developed for large commercial nuclear power plants \cite{lin2023generalization} \cite{li2024open}, as well as small modular reactors \cite{luo2024cross}. Stefenon et al. \cite{stefenon2024hypertuned} developed a hyper-tuned wavelet CNN with long short-term memory for time series forecasting in hydroelectric power plants. In the field of load prediction or forecasting, a multi-scale fusion CNN was presented in \cite{chen2024load} for load prediction of integrated energy systems. Further, Xu et al. \cite{xu2024framework} added attention mechanism in a time series depthwise separable CNN in a framework for electricity load forecasting. The CNN model offers significant benefits such as the preservation of spatial information, hierarchical feature learning, and translational invariance \cite{Goodfellow2016}. In addition, these networks are more efficient (requiring fewer parameters) compared to a similarly-tasked fully-connected network. These characteristics make CNNs superior when dealing with data exhibiting spatial correlations, making them appropriate to use with the 3D pin powers.

Acknowledging the existence of inherent uncertainties, the novelty of this work is to provide insight into the confidence levels of the CNN ML model's predictions through an UQ method, when the training dataset is relatively limited in size. Specifically, we will employ the MCD-based UQ technique, known for its effectiveness in quantifying uncertainties of the ML models.

\section{Methods} \label{sec:methods}

\subsection{General Approach} \label{sec:general_methods}

\begin{figure}[ht!]
    \centering
    \includegraphics[width=0.55\linewidth]{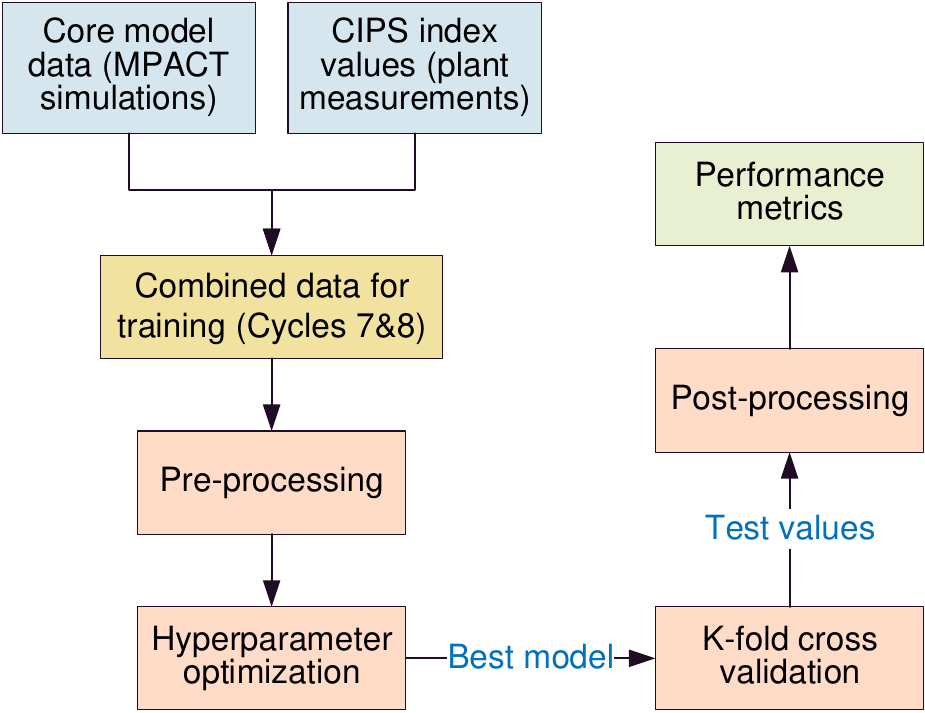}
    \caption{General workflow.}
    \label{fig:general_workflow}
\end{figure}

To illustrate the general process of this study, a visualization of the workflow is provided in Figure \ref{fig:general_workflow}. The core model data, generated using MPACT, are the inputs to the network and consist of the 3D pin powers, core exposures, and boron concentrations. For each of these entries, the corresponding CIPS Index calculated based on plant measurement data is then added. These arrays for Cycles $7$ and $8$ are then combined, including relevant information to identify the entries later on, such as assembly ID and cycle number. During pre-processing, the CIPS classes were computed by setting a hard threshold: an index value above 10 was classified as positive, and 10 or below as negative. This resulted in a binary-encoded vector of positive and negative labels. Each input channel (pin powers, core exposures, and boron concentrations) was independently scaled using min-max scaling (range: [0, 1]), shuffled, and partitioned using an $80$\%/$10$\%/$10$\% split for training, validation, and testing of the ML model. Scaling ensures that the model does not develop biases due to differences in magnitude, and shuffling prevents bias due to the ordering of instances. This data is then fed into the hyperparameter optimization algorithm, which trains a set number of models with unique hyperparameter configurations to minimize the validation loss. Once this process was complete, the best-performing model's hyperparameters were set for use in $k$-fold cross validation (CV).

The $k$-fold CV procedure involves taking the original dataset and re-splitting it $k$ times in terms of data partitioned for training and testing. It is important to note that the validation dataset used in the tuning process is excluded from this, as these data have already been used to optimize the hyperparameters. The hyperparameters fit using the validation dataset have the hypothesized potential to transfer knowledge of it, potentially leading to over-optimistic metrics if these data are not excluded \cite{boulesteix2015ten}. In each of the folds, the data for training and testing are uniquely split without replacement so that every available data entry is eventually used as a training instance and testing instance. Since each of these folds trains on a different training set, there are $k$ unique models at the end of the CV process. Assessing the differences in performance between these models can yield insight into how well the methodology generalizes across the entire (minus the validation set) dataset, in addition to being a valuable method for detecting overfitting. If all of these models perform similarly, it indicates that no particular portion of the dataset was favored.

Once the CV process is complete, each of the $k$ test values are exported for post-processing. Each of these test sets still represent true ``holdout'' sets, as these data were never seen during the training process in their respective folds or during hyperparameter tuning. As such, they can be combined to yield an estimate covering the entire dataset that was used for $k$-fold CV. This method yields more representative metrics when compared to the averaging of fold-computed metrics, which has the potential to produce biased values when significant class imbalances exist \cite{forman2010apples}. In post-processing, the array of test outputs and their corresponding inputs are scaled to their original ranges, separated by cycle number, and ordered by core exposure.

The prevention and detection of overfitting, defined as a decline in a model’s ability to generalize to new data despite high performance on training data, is an important concern in this study. The use of two cycles, totaling 2352 input sets with a class imbalance (15.2\% of input pairs qualify as positive CIPS), poses a potential risk of overfitting. To mitigate this, careful architecture design and hyperparameter tuning, as described above, balance model complexity to ensure the necessary features are learned without memorizing the input data. Exponential learning rate decay is employed to gradually reduce weight update rates, ensuring smooth convergence to the optimal solution. Training and validation losses are monitored to detect discrepancies that may indicate overfitting. Finally, the $k$-fold CV methodology ensures that the model generalizes well across all sections of the data, a strong indicator of strong performance on unseen datasets.

\subsection{Network Architecture} \label{sec:network_architecture}

The three input features consist of the core exposures, boron concentrations, and the 3D pin powers. The first two can be represented as vectors with lengths of $[2352]$, but the 3D pin powers have a shape of $[2352,17,17,44]$ with presumably spatially-sensitive relationships in terms of its relationship to CIPS. With this consideration, a 3D convolutional approach was chosen due to its ability to rapidly process spatial data without sacrificing arrangement and spatial relationships, among the other benefits discussed in Section \ref{sec:nn_background}. The network architecture uses a combination of two fully-connected stacks and a CNN to appropriately handle the shapes and physical characteristics of these input parameters. 

\begin{figure}[ht!]
    \centering
    \includegraphics[width=0.95\linewidth]{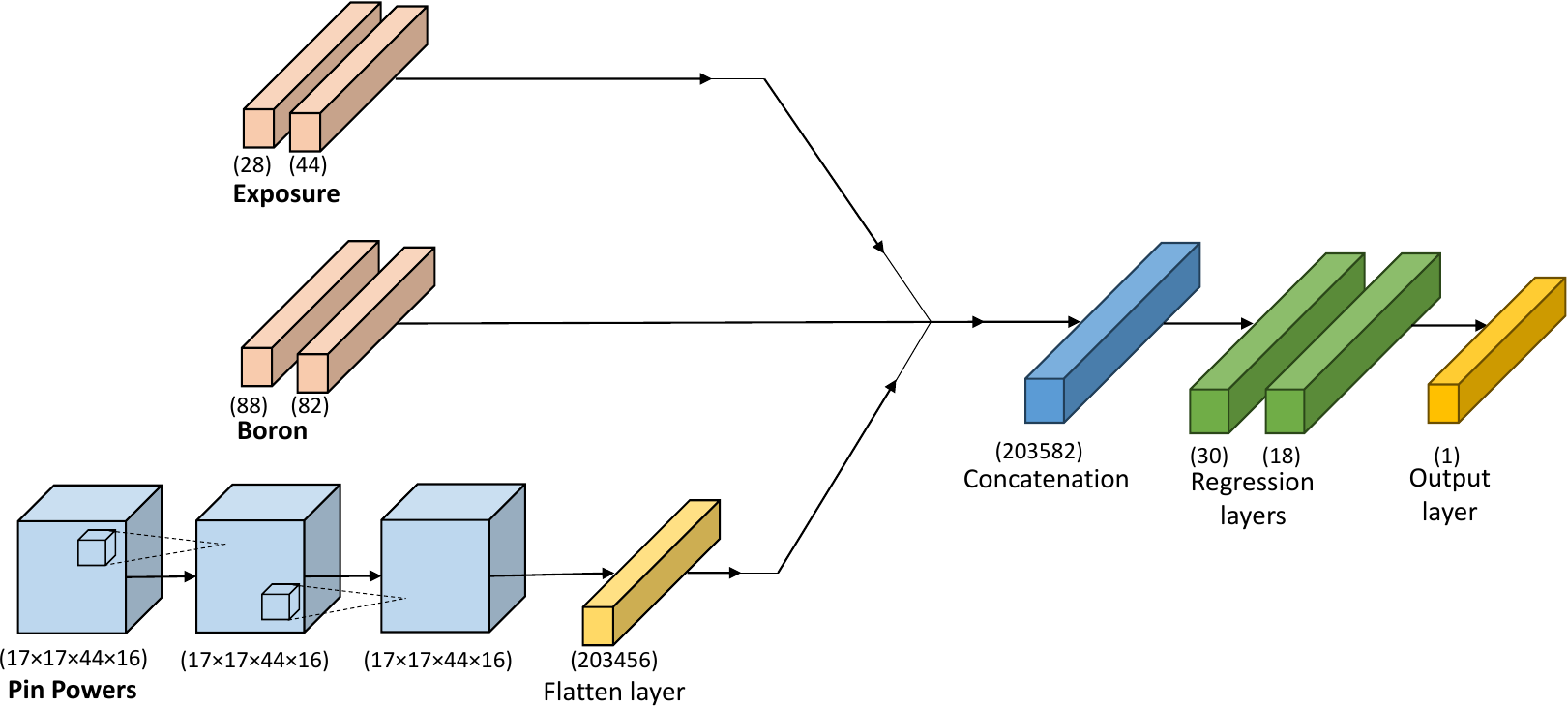}
    \caption{Visualization of the ANN architecture. Values below each component indicate the dimensions of each layer, such as the number of neurons in fully-connected stacks (e.g., boron and exposure). In the 3D convolutional layers, the three spatial dimensions are first indicated, followed by the number of filters. The dimensions of the flatten and concatenation layers are interpreted as the length of their output vectors. The values in this figure describe how a single set of inputs are processed, and do not indicate the volume of input data.}
    \label{fig:ann_architecture}
\end{figure}

Implemented using Google's TensorFlow \cite{abadi2016tensorflow}, the network architecture shown in Figure \ref{fig:ann_architecture} has a structure of three input channels making up the feature extraction section, which is followed by the classifier portion consisting of regression and output. The two scalar inputs (core exposure and boron concentration) are located in separate channels, each with two fully-connected (or densely-connected) layers. While these stacks could be combined into a single channel with further data preparation, this was empirically shown to decrease performance when compared to the presented architecture. The channel for the pin powers is made up of three 3D convolutional layers followed by a flattening layer to reduce the dimensions prior to regression. No pooling layers were used in this channel specifically to avoid creating translational invariance, where the absolute position of the feature in the ``image'' does not affect the prediction. While this is beneficial in other computer vision applications, the dependency that CIPS has with specific pin power features is not fully known. Because of this, there could be effects that depend on absolute position in the fuel bundle, such as the axial location of heightened pin powers. This comes with a trade-off in computational performance, as pooling would reduce the dimensionality between convolutional layers, but as discussed in Section \ref{sec:results}, this model is still relatively light-weight and this added expense is of little impact. For similar motivations, ``same'' padding was used in each of the 3D convolutional layers to maintain the input's dimensions throughout the stack. Without padding, information at the borders of the original ``image'' can be lost, as the output will be naturally smaller than the input. The padding process inserts zeros around the image, which effectively prevents this from occurring.

The flattened pin power features are then concatenated with the outputs from the exposure and boron channels, this vector then being passed into two fully-connected regression layers followed by the output layer. This output layer consists of a single neuron using a sigmoid activation function to produce a probability of instance membership to the positive class. In classifiers that predict for more than two classes, softmax would be a beneficial activation function, but for a binary classifier the sigmoid function is appropriate. The number of layers specified for the input channels, as well as the regression stack, were determined empirically while minimizing the validation loss using data from the validation partition prior to hyperparameter optimization. This is a commonplace approach, as there are no analytical means of optimizing network depth; the recommended technique is to increase depth until the validation loss fails to decrease further \cite{neuralnetdesign}. The dimensions of each of these layers are also provided in Figure \ref{fig:ann_architecture}.

Optimizing the values of hyperparameters is an important part of maximizing the performance of a classifier. The process of hyperparameter tuning was implemented for this architecture using RayTune, a package designed to automate tuning \cite{moritz2018ray}. A hyperparameter search space was first constructed with user-set boundaries for each of the $22$ hyperparameters. To adequately cover this search space, $100$ models were trained with unique hyperparameter configurations. The Asynchronous Successive Halving Algorithm (ASHA) was used to further optimize this process, which is an aggressive early-stopping algorithm designed to minimize resources spent on unsatisfactory configurations \cite{li2018massively}. This is accomplished by terminating hyperparameter configurations that report poor performance relative to their peers using a reduction factor. These models were allowed to train to a maximum of $100$ epochs.

\subsubsection{Performance Metrics}
\label{sec:performance_metrics}

To evaluate the performance of the classifier post-testing, the selection of appropriate metrics is important to gain insights into both its effectiveness and generalization capabilities. These measures serve as quantitative assessments of the model's predictive performance and the ability to discriminate between classes \cite{liu2014strategy}. The basis of these discrimination metrics \cite{hossin2015review} is in the tallying of positive and negative instances with respect to their classification correctness. Table \ref{tab:discriminator_metrics} provides a ``confusion matrix'' displaying each of these four tallies with additional context for the ease of the reader. 

\begin{table}[!htb]
	\centering
	\begin{tabular}{lcc}
		\hline
		& \textbf{Predicted Positive}  & \textbf{Predicted Negative} \\ \hline
		\textbf{Actual Positive}   & True Positive ($\text{TP}$) & False Negative ($\text{FN}$)\\
		\textbf{Actual Negative}   & False Positive ($\text{FP}$) & True Negative ($\text{TN}$)\\ \hline
	\end{tabular}	
	\caption{Discrimination-evaluation basis for a binary classifier.}
	\label{tab:discriminator_metrics}
\end{table}

Once these values are attained, the overall accuracy can be computed with Equation (\ref{eq:acc}), simply by taking the proportion of true classifications versus total number of classifications. Accuracy is one of the most widely used metrics due to its lack of complexity and ease of implementation. However, this metric alone has significant limitations in cases of class imbalance, where high accuracy values may be reported despite poor performance in the minority class \cite{japkowicz2013assessment}.

\begin{equation} \label{eq:acc}
    \text{accuracy} = \frac{\text{TP} + \text{TN}}{\text{TP} + \text{TN} + \text{FP} + \text{FN}}
\end{equation}

Another widely-used set of metrics are the true positive rate ($\text{TPR}$) and true negative rate ($\text{TNR}$), also known as the \textit{sensitivity} and \textit{specificity}. Sensitivity is interpreted as the probability of a positive classification, conditioned on the instance truly belonging to the positive class. Likewise, specificity is interpreted in the same manner but in terms of membership to the negative class. The $\text{TPR}$ and $\text{TNR}$ values can be computed using Equations (\ref{eq:tpr}) and (\ref{eq:tnr}). The false positive rate ($\text{FPR}$) and false negative rate ($\text{FNR}$) can also be computed by taking the complement of $\text{TPR}$ and $\text{TNR}$, respectively.

\begin{equation} \label{eq:tpr}
    \text{TPR} = \frac{\text{TP}}{\text{TP} + \text{FN}}
\end{equation}

\begin{equation} \label{eq:tnr}
    \text{TNR} = \frac{\text{TN}}{\text{TN} + \text{FP}}
\end{equation}

An additional metric employed is the $F_1$ score, which is defined as the harmonic mean between two measures known as \textit{precision} and \textit{recall}. The precision is the accuracy of the positive predictions made by the classifier, indicating how many instances were truly positive out of all positive predictions made. Recall is identical to the sensitivity ($\text{TPR}$) and can be interpreted as such. Precision, recall, and the $F_1$ metric are defined in Equations (\ref{eq:prec}), (\ref{eq:recall}), and (\ref{eq:f1}).

\begin{equation} \label{eq:prec}
    \text{precision} = \frac{\text{TP}}{\text{TP} + \text{FP}}
\end{equation}

\begin{equation} \label{eq:recall}
    \text{recall} = \frac{\text{TP}}{\text{TP} + \text{FN}}
\end{equation}

\begin{equation} \label{eq:f1}
    F_1 = \frac{2\cdot \text{precision} \cdot \text{recall}}{\text{precision} + \text{recall}}
\end{equation}

When considering the dataset used for this study, there is a clear and significant class imbalance between the CIPS-positive and CIPS-negative classes. The negative class has a majority of data instances, with just $15.2\%$ of all instances belonging to the positive class. In cases of significant class imbalance such as this one, the primary concern is that the minority class will not be adequately represented during training, potentially leading to poor predictive performance. Given this, the ability to evaluate the overall performance while considering the individual performance of both the majority and minority is essential. To provide a more balanced assessment in these cases, Flach et al. \cite{flach2015precision} proposes a modification to the existing $F_1$ score by weighting it based on the proportion of positive instances in the dataset, $\pi$. This adjusted metric is referred to as ``$F_1$ with gain'', or $FG_1$, and is computed with Equation (\ref{eq:f1g}).

\begin{equation} 
    \label{eq:f1g}
    FG_{1} = \frac{F_{1} - \pi}{(1 - \pi)F_{1}}
\end{equation}

Another useful metric in cases of class imbalance is the Matthews Correlation Coefficient (MCC). This metric was introduced in $1975$ by Brian Matthews and is a measure of association for two binary variables \cite{matthews1975comparison}. Outside of ML, it is known as the phi coefficient, which is identical to the Pearson Correlation Coefficient (PCC) estimated for two binary variables. The MCC allows for clear interpretation in the case of class imbalance, and is invariant to the swapping of classes \cite{chicco2020advantages}. The output values range from $[-1,1]$ with $-1$ indicating maximal disagreement between the predicted and reference labels and $+1$ indicating maximal agreement. To achieve a desirable coefficient value, desirable performance in both the majority and minority classes is required. As the MCC is a special case of the PCC, its interpretation is identical \cite{mukaka2012guide}. To compute the MCC in terms of a binary classifier, Equation (\ref{eq:mcc}) is used.

\begin{equation} 
    \label{eq:mcc}
    \textrm{MCC} = \frac{\text{TP}\times \text{TN}-\text{FP}\times \text{FN}}{\sqrt{(\text{TP}+\text{FP})(\text{TP}+\text{FN})(\text{TN}+\text{FP})(\text{TN}+\text{FN}})}
\end{equation}

\subsection{Uncertainty Quantification}
\label{sec:UQTheory}

\subsubsection{Overview of MCD for UQ of DNN}

The term UQ refers to the process of estimating the uncertainty associated with the predictions that ML models make. This process provides insights into the reliability of model predictions to inform on the level of ``trustworthiness'' that the model possesses. Quantifying uncertainty and reliability is critical in sensitive applications such as medical diagnosis, fraud detection, and autonomous driving. In terms of the application of this study, which is to be used in the core design process for PWR loading patterns, rigorous UQ methods must be implemented to address the reliability of this approach.

\begin{figure}[ht!]
    \centering
    \includegraphics[width=0.6\linewidth]{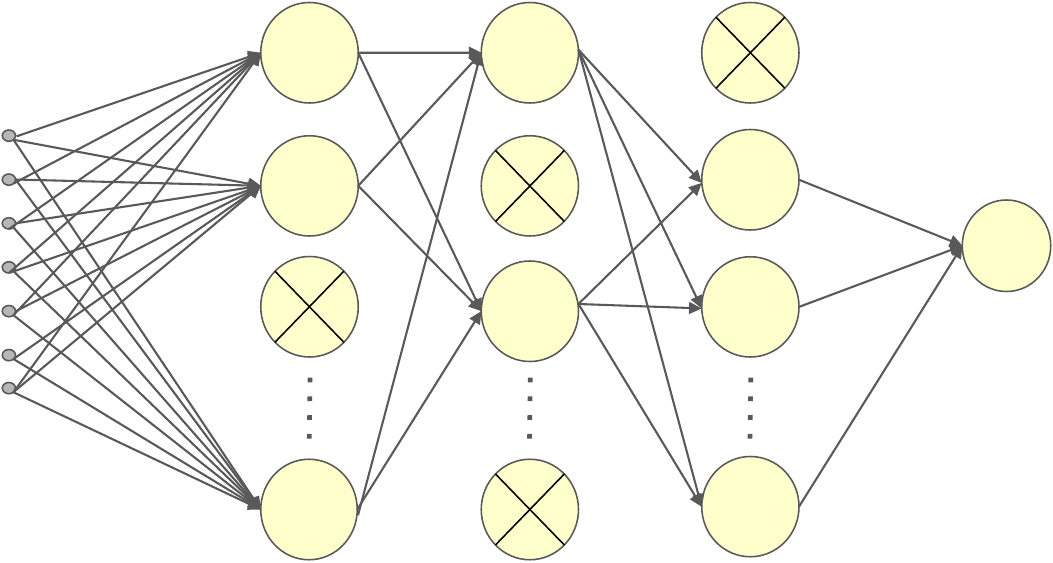}
    \caption{Visualization of active dropout layers in a simplified neural network.}
    \label{fig:dropout_visualized}
\end{figure}

Typical DNNs are known as ``black box'' models due to their inherent complexity, making clear interpretation difficult. The technique of MCD has emerged as an approach to help mitigate these concerns in the field of UQ as a Bayesian approximation \cite{gal2016dropout}. The basis of this approach is in leveraging the random nature of dropout layers in selectively disabling neurons for a given pass through the network, as shown in Figure \ref{fig:dropout_visualized}. Typically, these layers are only active during the training step, but in MCD they are allowed to be active in the prediction step as well after the training is completed. By introducing active dropout during both training \textit{and} prediction, MCD allows for an arbitrary number of predictions to be generated for a single input. The collection of predictions for a given input can be interpreted as random samples of the output distribution, permitting statistical analysis beyond what conventional DNN outputs are capable of. This technique also has the potential to reduce the risk of overfitting, particularly in cases of noisy or uncertain data \cite{goel2021robustness}. MCD, like all methods, is not immune from drawbacks. Its sensitivity to the dropout rate, which determines how many neurons are inactive, can greatly influence both a model's performance and uncertainty estimates \cite{seoh2020qualitative}. Higher rates may increase the regularization too much, leading to underfitting via excessive information loss, while rates too small can lead to overly confident and poorly calibrated uncertainties. Another drawback is the associated computational overhead, which can be considerable in resource-intensive settings, since inference requires multiple forward passes. MCD can also be less effective for tasks using recurrent neural networks, long short-term memory architectures, or structured outputs like image segmentation. However, for two-class problems, such as in this study, MCD is well-suited since the output probability can be modeled as a distribution with a mean and variance. Finally, to effectively use MCD in this context, defining appropriate uncertainty metrics is crucial for evaluating model performance \cite{kendall2017uncertainties}.

\subsubsection{Uncertainty Metrics}
\label{sec:mcd_metrics}

The implementation of MCD in this study is identical to that of the ``base'' model, with the use of $k$-fold CV. The key difference between the base model and MCD approach lies in the testing period, where instead of a set of single probability outputs, there are now $T$ probability outputs per input. For each of these input entries, the mean of the $T$ samples' probability values can be taken to provide an average predicted probability, $p^*$, using Equation (\ref{eq:mcd_means}).

\begin{equation} \label{eq:mcd_means}
    p^* = \frac{1}{T} \sum^T_{t=1} p_t
\end{equation}
where $T$ is the number of MCD evaluations.

A useful metric of uncertainty is the concept of \textbf{entropy}, which is a measure of the amount of information needed to describe the outcome of a random event. \textbf{Predictive entropy} is defined as the entropy of the mean of the predictive distribution, and is a value computed for each input entry using the mean probabilities for each class in a given entry \cite{camarasa2020quantitative}. This value reaches a maximum when the model is maximally uncertain with equal class probabilities of $0.5$ in the case of binary classification. The formulation of predictive entropy is shown in Equation (\ref{eq:pred_entropy}).

\begin{equation} \label{eq:pred_entropy}
    \mathbb{H} \approx -\sum^C_{j=1} p^*_j \log(p^*_j)
\end{equation}
where $C$ is the number of output classes.

In a specific model, there are two types of uncertainty: \textit{epistemic} and \textit{aleatoric}. The former, also known as systematic uncertainty, arises when there is an incomplete understanding or lack of data that prevents an effective model from being developed. This is reducible as better approaches or data are found. Aleatoric uncertainty is inherent to the data due to stochastic effects, and is irreducible as such \cite{smith2018understanding}. When assessing the uncertainty of a ML model, it is particularly desirable to quantify its epistemic uncertainty. While predictive (marginal) entropy is an effective measure of uncertainty, it only provides insight to the combined effects of epistemic and aleatoric origins and does not uniquely identify either. An approach to disseminate between the two can be found when considering \textbf{mutual information} ($\mathbb{I}$). This measure is given by the difference between the marginal entropy and the conditional entropy, computed in the context of MCD using Equation (\ref{eq:mi}).

\begin{equation} \label{eq:mi}
    \mathbb{I} \approx \mathbb{H} - \sum^C_{j=1}\frac{1}{T} \sum^{T}_{t=1} p_{jt} \log(p_{jt})
\end{equation}

Specifically, this relation quantifies the amount of information gained about model parameters when given a label for a new input point. If the model parameters are well-determined, a small amount of information would be gained. On the other hand, high uncertainty about an input point suggests that knowledge of the label would yield a high information gain \cite{smith2018understanding}. The interpretation of this is that for mutual information between the true label and the model parameters, a high $\mathbb{I}$ value would indicate a larger uncertainty for a given prediction, while a low $\mathbb{I}$ value would indicate a smaller uncertainty.

Another insightful uncertainty metric is the \textbf{margin of confidence} ($\mathbb{M}$), which is the difference in the two most confident predictions in each class \cite{milanes2021monte}\cite{guochen2021four}. In terms of this study, this is defined as the difference between the maximum predicted probability of the positive class and the maximum predicted probability of the negative class over the $T$ samples for a given input. This value can be computed with Equation (\ref{eq:moc}) with larger values suggesting that the model is more confident in its prediction, as the highest predicted probability is substantially higher than that of the opposing class. Conversely, low values of $\mathbb{M}$ indicate that the model is less certain with closer probabilities in opposing classes. This metric carries a maximum value of $1.0$ and has the potential to be negative.

\begin{equation} \label{eq:moc}
\begin{aligned}
    c &= \argmax_j p^*_j  \\
    d_t &= p_{ct}-\max_{\substack{j \neq c}} p_{jt}  \\
    \mathbb{M} &= \frac{1}{T}\sum^T_{t=1}d_t
\end{aligned}
\end{equation}

With the method of MCD and relevant metrics to quantify the outputs, it is now possible to accurately assess the level of uncertainty that the trained classifier possesses. This essential UQ step will allow for a determination of the trustworthiness and reliability of the model \cite{kendall2017uncertainties}.

\section{Results} \label{sec:results}

\subsection{Results for Base Model} \label{subsec:timing}

Each of the five folds was trained over a period of $100$ epochs using their unique training datasets. On average, each fold took $43.74$ minutes to train using four Apple M1 Pro cores in resources. The five folds, which are uniquely-weighted networks, were then frozen and used to predict against their test sets. The combined inference time of these folds was $0.0172$ seconds to essentially predict two cycles worth of data. The breakdown of these time expenses is located in Table \ref{tab:base_time_stats}. 

The lightweight nature of the trained model is a significant advantage. By comparison, the current coupled CTF/MAMBA methodology, integrated with the SIMULATE core simulator, typically requires one to two hours of runtime using 193 cores \cite{andersen2021machine}. Although the MAMBA approach offers higher-fidelity outputs, such as pinwise crud mass projections, the neural network in this work is designed for the rapid prototyping of core loading patterns. Finalized designs would then be evaluated using the traditional methods that are currently in use. With entire-core, full-cycle prediction times under 20 milliseconds, the model could be readily incorporated into existing drag-and-place graphical user interface tools for first-order LP prototyping.

The total training time of $215.9$ minutes using limited resources is unlikely to hinder deployment. This method's small computational needs would allow the model to be re-trained quickly when new data becomes available, building a more effective model over time. Overall, these time savings when compared to traditional physics-based models make it a viable tool to improve workflow efficiency in the core design process.

\begin{table}[!htb]
	\centering
	\begin{tabular}{lcc}
		\hline
                & \textbf{Training (min)} & \textbf{Testing (s)} \\ \hline
		\textbf{Fold 1} & $48.95$ & $\num{3.64e-3}$ \\
		\textbf{Fold 2} & $43.60$ & $\num{3.66e-3}$ \\
        \textbf{Fold 3} & $43.43$ & $\num{3.55e-3}$ \\
		\textbf{Fold 4} & $43.45$ & $\num{3.15e-3}$ \\
        \textbf{Fold 5} & $36.46$ & $\num{3.21e-3}$ \\ \hline
        \textbf{Total}  & $215.88$  & $\num{1.72e-2}$ \\ \hline
	\end{tabular}	
	\caption{Time expenditure over the folds.}
	\label{tab:base_time_stats}
\end{table}

\begin{figure}[ht!]
    \centering
    \includegraphics[width=0.7\linewidth]{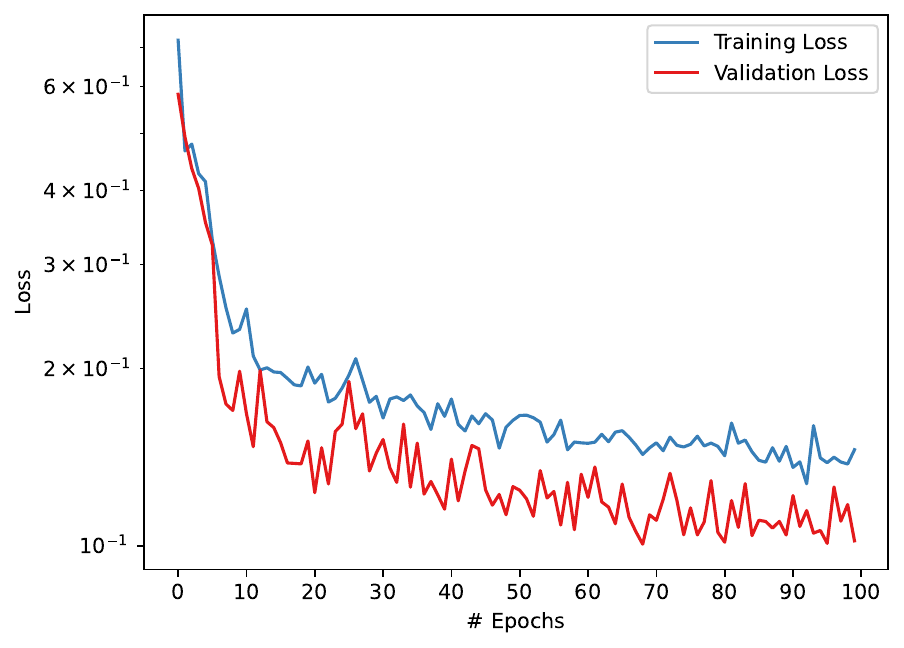}
    \caption{Training and validation loss curves for fold $1$. The other folds showed similar behavior.}
    \label{fig:training_loss_curves}
\end{figure}

Inspection of the training and validation loss curves can be a useful tool in identifying potential overfitting. For this model, these curves are shown for Fold $1$ over the training period in Figure \ref{fig:training_loss_curves}. The validation loss is shown below the training loss at every epoch, an effect of using dropout layers within the architecture. The dropout layers are active when training over a given epoch, but inactive when evaluating the validation set, which can lead to inflated recorded training losses. Both curves follow a similar trend over the $100$ epochs, with a majority of the loss reduction occurring over epochs one to eight, and do not indicate the presence of overfitting. Validation loss did not significantly improve after $70$ epochs, and the model was observed to be stable out to the training period's conclusion at $100$ epochs. The curves for each of the other four folds were also inspected, with the same trends observed.

\subsubsection{Base Metrics}
\label{sec:base_metrics}

Once the $k$-fold CV was completed, each of the fold's prediction results (true positive, false positive, etc.) were then concatenated into a single set for metric evaluation. This approach combines each of the folds' test results, and since the $k$-fold without replacement procedure enforces that every entry of the original dataset (with the validation set omitted) is eventually seen as a test point, a vector is produced effectively reconstructing the entire dataset as test entries. This method is preferred over the traditional averaging of fold-computed metrics, which can introduce bias, particularly in cases of a large class imbalance with a small proportion of positive class members \cite{forman2010apples}. This practice does not affect the accuracy or error rates, which will be the same when averaged or taken from the concatenation, but provides a less biased assessment of the $F_1$, $FG_1$, and MCC values. A listing of the selected metrics is included in Table \ref{tab:concatenated_metrics}, with an overall accuracy of $92.3$\%. Despite its drawbacks in the case of this study, the $F_1$ score is also reported due to its widespread use in classification problems.

\begin{table}[!htb]
	\centering
	\begin{tabular}{lc}
		\hline
		\textbf{Metrics}       & \textbf{Values} \\ \hline
		Accuracy              & $0.923$  \\
		True Positive rate    & $0.832$ \\
        False Positive rate   & $0.056$ \\
		True Negative rate    & $0.944$ \\
        False Negative rate   & $0.168$ \\
        $F_1$ score  & $0.798$ \\
        $FG_1$ score & $0.939$ \\
        MCC                   & $0.752$ \\ \hline
	\end{tabular}	
	\caption{Metrics of concatenated $k$-fold test predictions.}
	\label{tab:concatenated_metrics}
\end{table}

It is difficult to interpret the classification performance by using any one metric alone \cite{chase2014thresholding}, 
especially in cases of class imbalance. However, assessing the classifier's performance over several metrics can provide valuable insight. The true positive rate of this classifier, $83.2$\%, is lower than the true negative rate of $94.4$\%, indicating that the classifier does have some bias towards negative class instances. This is not entirely unexpected due to the class imbalance within the training set. Yet, with an MCC value of $0.752$, the predictions can be interpreted as having a ``high positive correlation'' to the true class labels \cite{mukaka2012guide}. The $FG_1$ score has a value of $0.939$, further indicating that the classifier generally performs well in predicting both positive and negative classes.

The test predictions from each of the folds were separated and then used to compute metrics on a fold-wise basis. These metrics are compared in Table \ref{tab:fold_wise_performance} using four selected metrics: accuracy, $F_1$-score, $FG_1$-score, and the MCC. In terms of accuracy, each of the folds perform similarly, with a maximum difference of $0.01$ compared to the computed mean. Larger amounts of variance are observed in the fold-wise $F_1$ scores, with the best-performing fold having a difference of $0.05$ compared to the lowest-performing. The re-weighted $FG_1$ metrics produced similar values over the five folds. The MCC distribution was observed with similar degree of variance when compared to the base $F_1$ scores, with an equivalent maximum difference of $0.05$ between folds. This variation could suggest that some of the folds had more informative training data, which could be explained by the random selection process for each fold's training sets. With the limited nature of this effect, coupled with the even performance in accuracy and $FG_1$ scores, this surrogate model adequately generalizes across the original dataset without significant preference for any single region. 

\begin{table}[!htb]
	\centering
	\begin{tabular}{lcccc}
		\hline
                & \textbf{Accuracy} & \textbf{$\mathbf{F_1}$} & \textbf{$\mathbf{FG_1}$} & \textbf{MCC} \\ \hline
		\textbf{Fold 1} & $0.929$ & $0.778$ & $0.950$ & $0.739$ \\
		\textbf{Fold 2} & $0.931$ & $0.828$ & $0.952$ & $0.787$ \\
        \textbf{Fold 3} & $0.921$ & $0.792$ & $0.937$ & $0.743$ \\
		\textbf{Fold 4} & $0.923$ & $0.779$ & $0.935$ & $0.736$ \\
        \textbf{Fold 5} & $0.913$ & $0.807$ & $0.942$ & $0.764$ \\ \hline
	\end{tabular}	
	\caption{Prediction metrics in each of the five folds.}
	\label{tab:fold_wise_performance}
\end{table}

\subsubsection{Exposure-Dependent Efficacy}
\label{sec:exposure_dependent_efficacy}

\begin{figure}[ht!]
    \centering
    \includegraphics[width=0.8\linewidth]{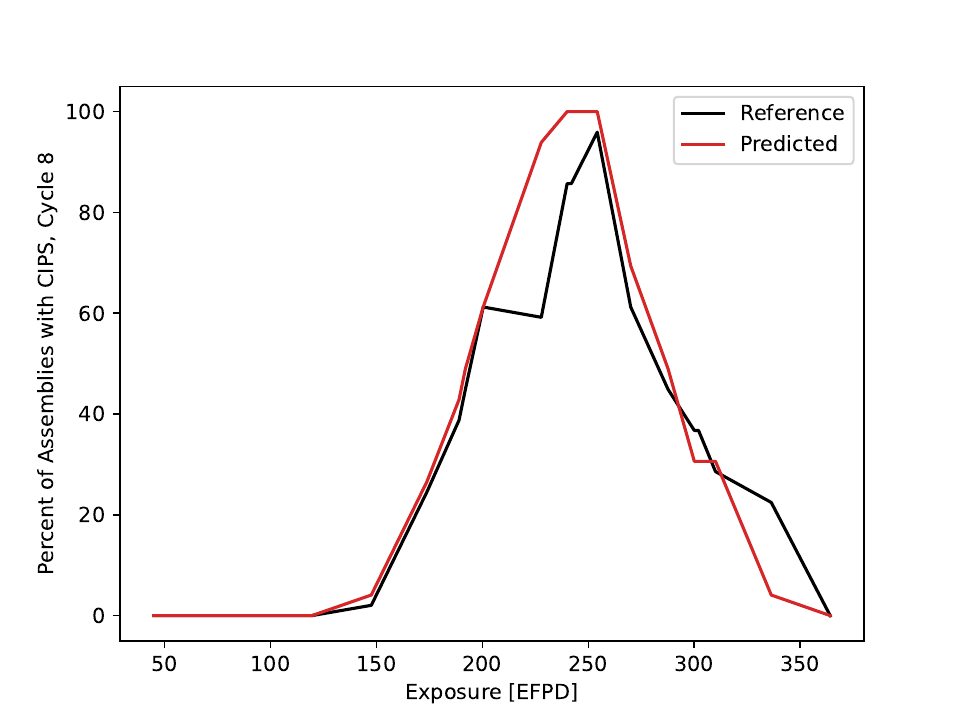}
    \caption{Proportion of CIPS-affected assemblies for predicted and true values in Cycle 8.}
    \label{fig:percent_cips}
\end{figure}

The concatenated predictions were then partitioned based on cycle number and sorted by exposure within these partitions. At each of these $24$ exposures, the predicted number of CIPS-positive instances were compared to the CIPS-negative instances. This yields a proportion of total assemblies that are predicted to have CIPS, expressed as a percentage at each exposure step over the cycle. The same process was then applied to the reference values, and both of these curves then plotted in Figure \ref{fig:percent_cips} for Cycle 8. The predicted curve agrees completely with the reference until after $119$ EFPD, when the onset of CIPS occurs. The rise in CIPS instances after this point is still well-approximated by the predicted curve, with a maximum deviation of $4.1\%$ until $200$ EFPD, where a plateau occurs in the reference set. This plateau is out-of-trend and without a clear explanation, as there were no significant changes in plant operating parameters in the provided data. The predicted values do not adhere to this unexpected dip and continue to rise with a decreasing slope until reaching a maximum at $240$ EFPD. This maximum predicted value remains at $100$\% until $254$ EFPD. At this same exposure, the reference curve reaches its highest magnitude with $95.2\%$ of assemblies experiencing CIPS. After this point, both predicted and reference curves are observed decreasing at similar rates until $300$ EFPD, where the predicted curve is seen under-predicting CIPS prevalence for the remainder of the cycle before reaching zero at $364$ EFPD.

\begin{figure}[ht!]
    \centering
    \includegraphics[width=0.8\linewidth]{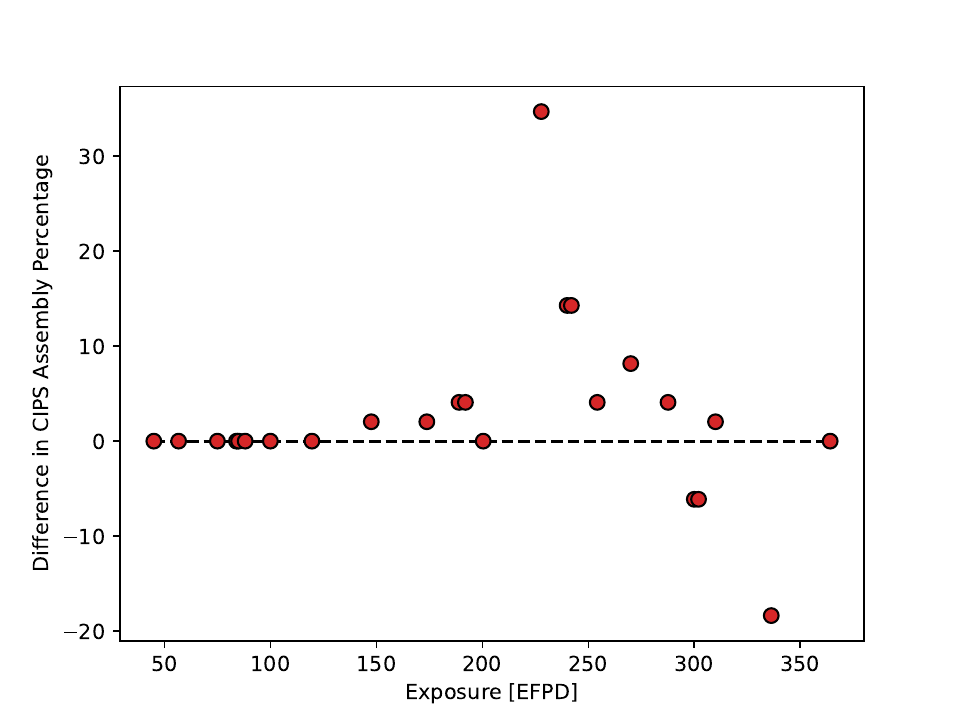}
    \caption{Residuals for assembly percentages over the course of Cycle 8.}
    \label{fig:percent_residuals}
\end{figure}

To illustrate the trends in agreement between these two curves, the residuals were plotted in Figure \ref{fig:percent_residuals}. These points were computed by taking the difference between the predicted and reference percentage values at each exposure step. As seen in Figure \ref{fig:percent_cips}, there is negligible deviation between the surrogate model's predictions and the true values in the early exposures. At the point of CIPS onset, after $120$ EFPD, there is a modest rise in this deviation that abruptly increases to a maximum value at $227$ EFPD. In Figure \ref{fig:percent_cips}, this corresponds to the plateau observed in the reference-computed curve. At the peak of CIPS instances, at $254$ EFPD, the deviation between measured and predicted values is $4.08\%$. The remainder of Cycle 8 sees agreement within $\pm 10\%$ except for a late-cycle outlier at $336$ EFPD. The placement of these residuals do not indicate an independent or identically distribution.

Even with the central outlier removed, there does appear to be some degree of autocorrelation in the mid-cycle range. This suggests the existence of a systematic bias in the model's predictions, particularly in the case of the CIPS-positive region. This trend may indicate a limitation in the neural network's architecture or an inadequacy in the representation of underlying relationships. The limited and imbalanced nature of the training set could also contribute to this, as too few positive training points can inhibit generalization to these areas where positive instances are prevalent. Despite this effect, out of the $24$ exposures, $20$ of these residual points lie between $\pm 10\%$ of the reference values.

With the above analysis in an exposure-dependent frame, this surrogate model can effectively predict when the onset of CIPS occurs, when the peak occurs and at what magnitude, and can capture the overall trajectory of CIPS development over Cycle 8. There are two regions of deviation greater than $10\%$, and the exact causes of the reference curve's out of trend values are not fully understood. Regardless, these results indicate that the surrogate model does capture the temporal effects of CIPS formation and resolution for this cycle.

This methodology was then applied for the two cycles that did not contain any CIPS occurrences, Cycle 6 and Cycle 7. These were completely ``clean'' cycles that can serve to assess the models' potential tendencies for false positives. Since Cycle 6 was not included in the training data, the trained Fold $1$ was used to directly predict the entire cycle's assemblies at each exposure. Since Cycle 7 was contained in the training data, the concatenated Cycle 7 test predictions from the $k$-fold CV were used. All but four of Cycle 6's predicted values were true negatives, with false positives accounting for $0.48\%$ of all predictions made for this cycle. Concatenated Cycle 7's test predictions saw zero incorrect predictions, with all of these being true negatives.

After assessing the performance of this surrogate modeling approach by plotting the proportions of CIPS-positive assemblies over time, a more practical perspective can be taken. If deployed in a core design environment, these exposure-related predictions would be relevant information that an engineer could use for a prospective loading pattern. Between the numerical performance metrics in addition to the plotted trajectories, this model can accurately predict the positive instances of CIPS in affected cycles while avoiding false positives in CIPS-negative cycles.

\subsection{Results for Uncertainty Quantification}
\label{sec:UQResults}

To implement MCD, the same methodology utilizing $k$-fold CV was maintained from Section \ref{sec:general_methods}. The architecture was modified to apply dropout on all hidden layers, to ensure that uncertainty is propagated throughout the network. It is of note that three-dimensional spatial dropout was applied to the convolutional elements of the model, as traditional dropout can remove features that may be spatially important \cite{ghiasi2018dropblock}. The training process was identical to that of the base model, but once the network was trained, the predictions made on the test cases were subject to MCD. This was performed by setting the model to prediction-mode while also allowing it to be trainable. The weights of this model are still frozen, but it allows for the dropout layer parameters to vary stochastically. The impact of this is that when the same input enters the neural network model multiple times, different outputs will be obtained since different neurons are dropped each time, the ensemble of which can represent the uncertainty in the output.

\subsubsection{MCD Prediction Performance}

Over the course of the $5$ folds, each of the test inputs was predicted for a total of $T = 50$ times to produce $50$ distinct output values while using a dropout rate of $0.1$. This value was empirically optimized using a grid search with the validation partition, with the objective to maximize prediction performance while maintaining reasonable uncertainty estimates (evaluated via their calibration). For each of these sets of $50$, the means and variances were then taken for an assessment of their uncertainties. The means were then treated in the same manner as the base model's output probabilities, and were split at a threshold of $0.5$ to determine the predicted CIPS class. With the vector of class predictions and the reference values, the conventional and class-balanced metrics were then applied and reported in Table \ref{tab:concatenated_metrics_mcd}. When compared to the performance results of Section \ref{sec:base_metrics}, the MCD results achieve the same level of accuracy while exhibiting slightly reduced performance in other metrics. The differences in these metrics with those of the base model are less than $0.01$ in all cases. Although the technique of MCD does not increase the performance of this classifier within these metrics, the close agreement of these values and those of the base model reflects the classifier's ability to consistently and repeatably make predictions \cite{lemay2022improving}.

\begin{table}[!htb]
	\centering
	\begin{tabular}{lc}
		\hline
		\textbf{Metrics}       & \textbf{Values} \\ \hline
		Accuracy              & $0.923$  \\
		True Positive rate    & $0.824$ \\
        False Positive rate   & $0.055$ \\
		True Negative rate    & $0.945$ \\
        False Negative rate   & $0.176$ \\
        $F_1$ score  & $0.796$ \\
        $FG_1$ score & $0.938$ \\
        MCC                   & $0.749$ \\ \hline
	\end{tabular}	
	\caption{Metrics of concatenated test predictions using mean MCD values.}
	\label{tab:concatenated_metrics_mcd}
\end{table}

\subsubsection{Model Calibration}

The term \textit{calibration} in this work refers to the state when a model's predicted probabilities correspond to the true occurrence in the reference class. It should not be confused with the calibration of a physical model such that its prediction matches the observation. Even with the high degree of efficacy presented in the previous metrics, it is still necessary to assess how the model's probability estimates are calibrated \cite{xenopoulos2022calibrate}. Reliability diagrams are useful tools in assessing the performance of a binary classifier, serving as a graphical representation of the alignment between predicted probabilities and true outcomes. The predicted probabilities are first partitioned into discrete bins, and then the reference labels are collected and aggregated within each of these probability bins, with the proportion of positives computed \cite{niculescu2005predicting}. \textit{Perfect calibration} would mean that this proportion of positives is equal to the bin's mean probability. For example, an ideal classifier's bin with a mean probability of $0.6$ should have a set of reference labels that are $60$\% made up of positive members. 

\begin{figure}[ht!]
    \centering
    \includegraphics[width=0.8\linewidth]{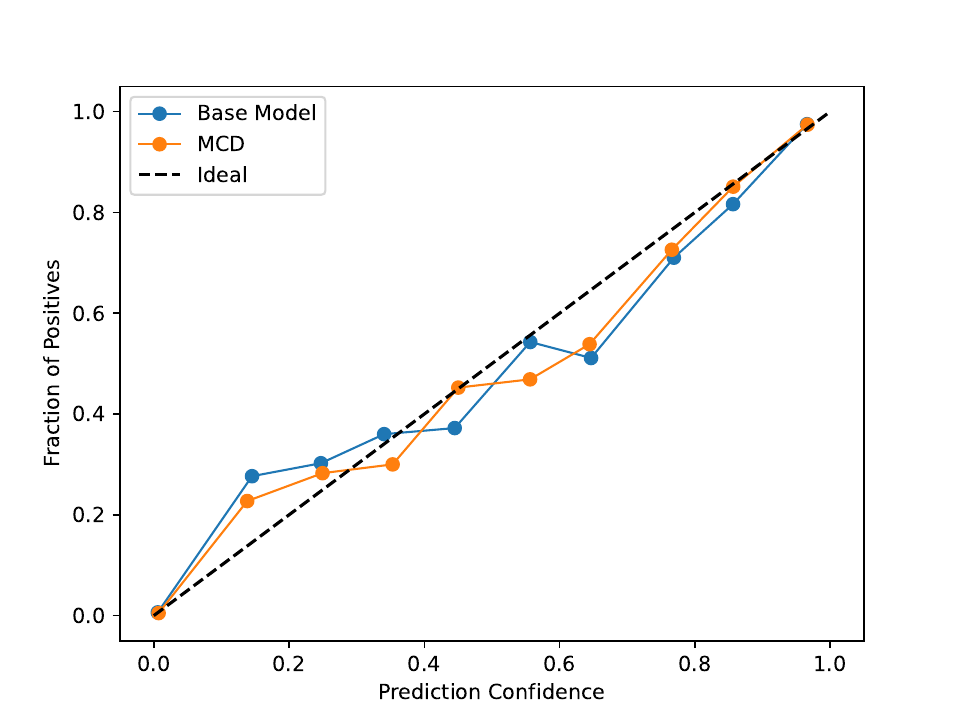}
    \caption{Reliability diagrams comparing base model and MCD.}
    \label{fig:mcd_calibration}
\end{figure}

This method was applied to the ``base model'' CIPS classifier as well as the MCD model and plotted together with 10 bins in Figure \ref{fig:mcd_calibration}. Most bin points for both models are seen relatively close to the ideal calibration line, with more deviation noted at regions between the center and each terminus. Where the mean bin probabilities are lower, there are a disproportionate fraction of positives, which would be indicative of the model slightly ``under-forecasting''. This is interpreted as the model is being conservative with predicted probabilities around this region. On the other hand, mean bin confidences around $0.6$ see the opposite, where bins $7$ through $9$ are below the identity line at different magnitudes. This indicates that the classifier is ``over-forecasting'' and overconfident in these bins. These are samples which were predicted to have a higher probability than the true rate, an undesirable quality. Both of these regions of over and under-confidence are limited in magnitude.

In all but two bins, the MCD values are closer to the identity line compared to those of the base model. Overall, the trend of these points are similar to the base model, with under-confidence observed in the lower probability bins and over-confidence in predictions for the upper probability bins. These effects are less pronounced compared to the base model, indicating that the MCD technique can lead to an increased overall calibration.

As a quantitative assessment of calibration, the non-MCD model achieved a Brier score of 0.054, while the MCD model reported a slightly improved score of 0.052. The Brier score, which measures the mean squared difference between predicted probabilities and actual outcomes, ranges from 0 (perfect calibration and accuracy) to 1 (completely miscalibrated or inaccurate predictions). Although the improvement in the Brier score appears modest, it reflects a measurable improvement in the MCD model’s ability to provide more accurate and calibrated probability estimates. This reduction indicates that the MCD model better aligns its predicted probabilities with the true likelihood of outcomes, particularly in regions where the non-MCD model exhibited overconfidence or underconfidence in its predictions. When combined with visual evidence from the reliability diagrams, these results suggest that MCD not only improves calibration but also improves the overall reliability of the model’s probabilistic outputs.

\subsubsection{Assembly-wise Confidence} \label{subsubsec:assemblywise_confidence}

The concatenated mean and standard deviation values from the MCD $k$-fold runs were then partitioned into their respective cycles and ordered by exposure and assembly ID. Taking into account an assumption of symmetric geometry, this leaves $38$ test predictions for unique assembly locations at each step in exposure. In the symmetric quarter-core for CNS-1, there are a total of $56$ assemblies. To populate the remaining $18$ locations for analysis, additional input sets were taken from the MPACT \texttt{c8.h5} file, with MCD predictions later performed on them. These assembly locations were omitted from the primary dataset due to the absence of instrumentation and therefore CIPS Index values. While this means that the predictions made for these assemblies cannot be validated, the MCD statistics can still be helpful to identify any trends over the quarter-core. To obtain these values, the same MCD methodology described above was applied with $50$ samples. With this considered, it is important to note that the quarter core assimilation may not appear to have the same CIPS percentage from Section \ref{sec:exposure_dependent_efficacy} as symmetrically-duplicate assemblies have been omitted from each map.

\begin{figure}[ht!]
    \centering
    \includegraphics[width=0.7\linewidth]{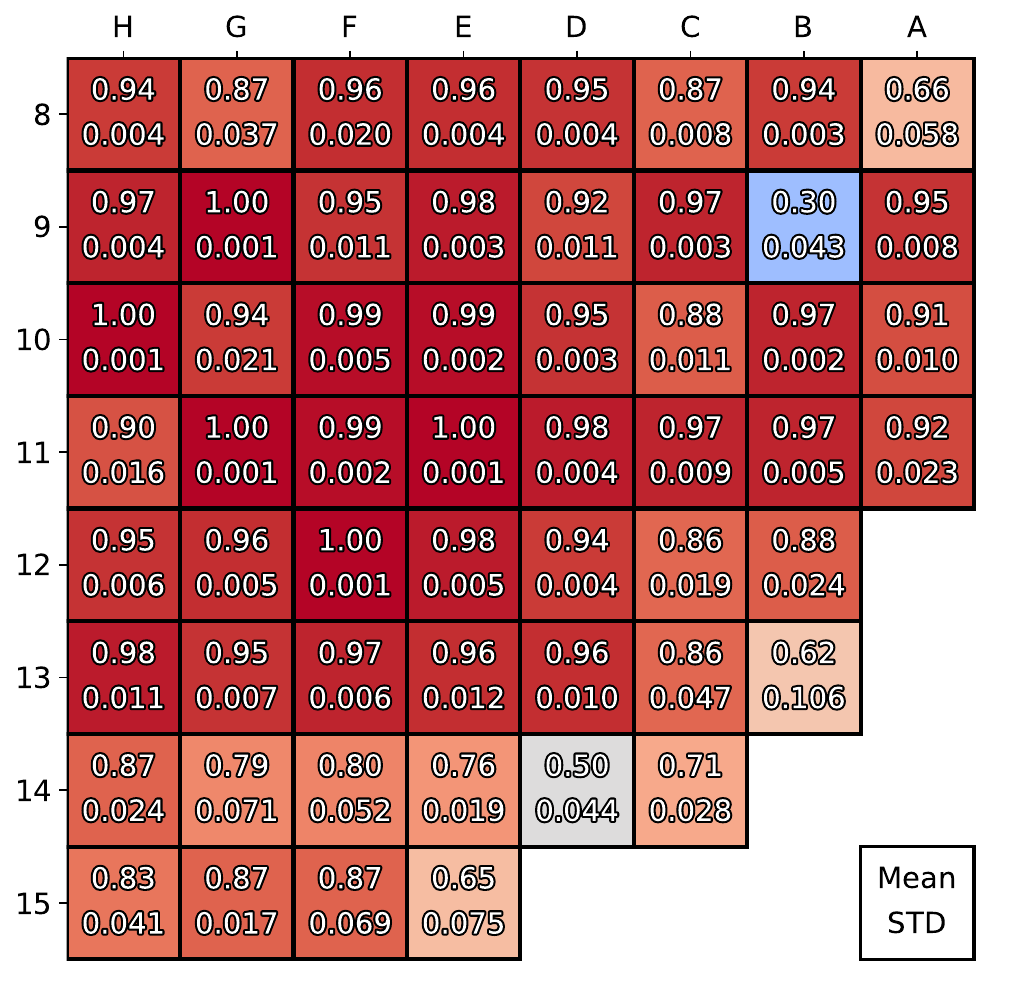}
    \caption{Mean and standard deviation values for each assembly prediction at 254 EFPD in Cycle 8.}
    \label{fig:quarter_254efpd}
\end{figure}

The first exposure considered is $254$ EFPD, when CIPS is the most prevalent, with $95.9\%$ of the instrument-loaded assemblies being affected. These assemblies and their respective mean MCD probabilities, along with their standard deviations, were then plotted as a plan view map in Figure \ref{fig:quarter_254efpd} to show the distribution of the classifier's confidences. With a classification threshold of $0.5$, mean values closer to $1$ indicate more confidence in a positive label, and values closer to $0$ indicate more confidence in a negative label. The heat map of this figure is determined by the mean value from light blue to red, corresponding to negative and positive predictions. 

The higher confidences populate the center of the quarter-core map, with smaller values occupying the far periphery. With a classifier threshold, one assembly (B-09) is predicted to be CIPS-negative. There is a trend of higher standard deviations being associated with more ``unsure'' predictions that are closer to the threshold of $0.5$. These locations are distributed more frequently near the bottom and rim of the quarter, with an outlier present on B-09 predicting a CIPS assembly with a higher confidence. The underlying reason for this outlier could be due to differences in the 3D pin powers, as the other input parameters (boron concentration and exposure) are spatially constant over this step. Overall, the mean of the predicted positive class' standard deviations is $0.0181$ making nearly every positive predictions' mean at least $3\sigma$ away from the threshold.

\begin{figure}[ht!]
    \centering
    \includegraphics[width=0.7\linewidth]{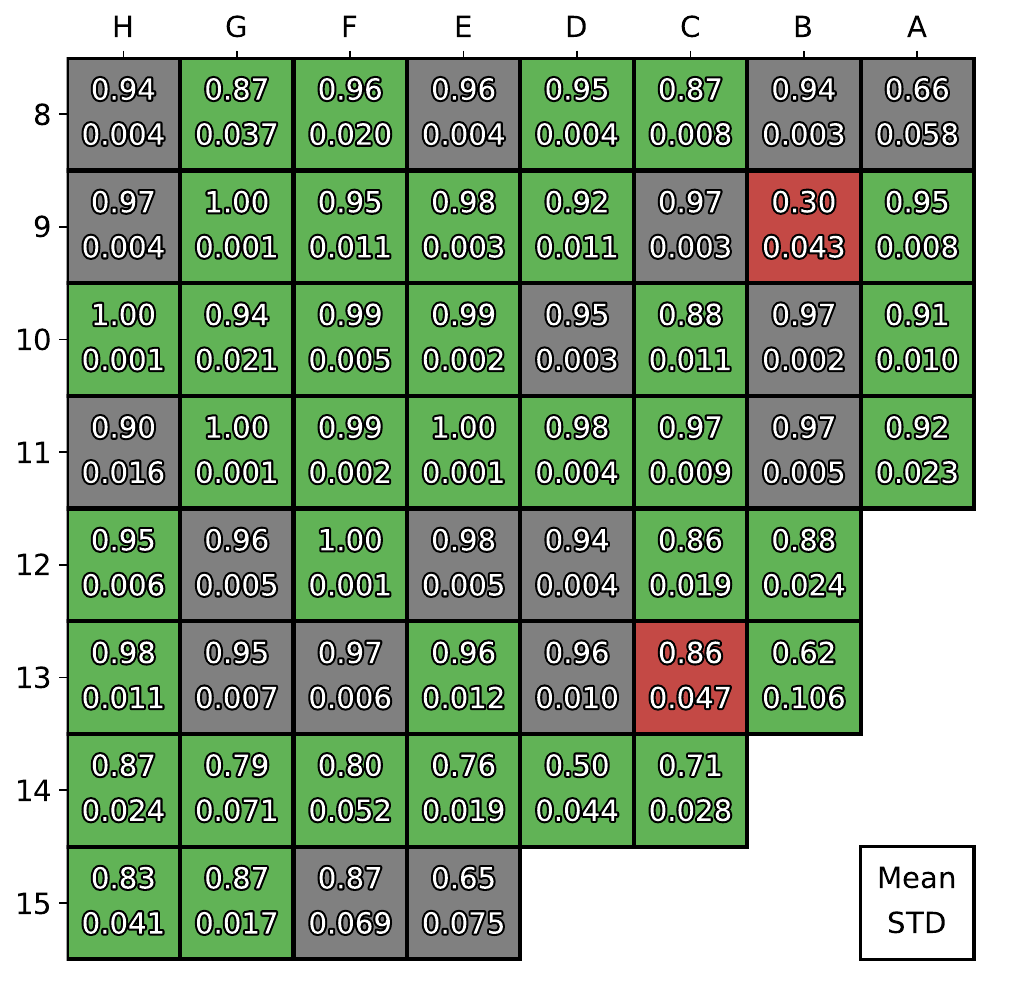}
    \caption{Mean and standard deviation values against prediction correctness for each assembly at 254 EFPD in Cycle 8.}
    \label{fig:quarter_254efpd_accuracy}
\end{figure}

The heat map was then replaced with binary colors in Figure \ref{fig:quarter_254efpd_accuracy} to indicate whether a correct or incorrect prediction was made. Green indicates that a given assembly's prediction was the same as the reference classification, while red corresponds to a prediction made in disagreement with the reference. The gray assemblies indicate that these predictions were made from the additional input entries without CIPS data to validate with. Both the means and the standard deviations were carried over from Figure \ref{fig:quarter_254efpd} to assess the relative confidence in correct or incorrect predictions.

There were $36$ correct predictions made out of a total of $38$ for this quarter core. The outlier previously identified, B-09, was one of these in addition to C-13 to the southeast. Despite a mean relatively higher to those in the surrounding region, C-13 was incorrectly classified as being CIPS positive. The true value, CIPS-negative, breaks the local trend where every neighboring assembly is CIPS-positive in addition to the larger region. Returning to the original dataset with CIPS Index values, this location is an edge case with a value within $0.5$ of the CIPS threshold of $10$. When considering the noise inherent to the dataset, this input set may have features indicative of CIPS but with an inaccurate labeling due to the nature of the CIPS Index metric. In the case of B-09, there does not appear to be a straightforward hypothesis for its existence, as this region is entirely made up of CIPS-positive assemblies.

\subsubsection{Bivariate Analysis}

One of the major benefits of using MCD is allowing for the treatment of predictions as statistical distributions. Implementing the metrics as described in Section \ref{sec:mcd_metrics}, the \textbf{margin of confidence} ($\mathbb{M}$) and \textbf{mutual information} ($\mathbb{I}$) values were computed for Cycle $8$'s predictions. For each set of MCD samples (i.e., for each input), a single value for $\mathbb{M}$ and $\mathbb{I}$ was calculated, respectively.

\begin{figure}[ht!]
    \centering
    \includegraphics[width=0.7\linewidth]{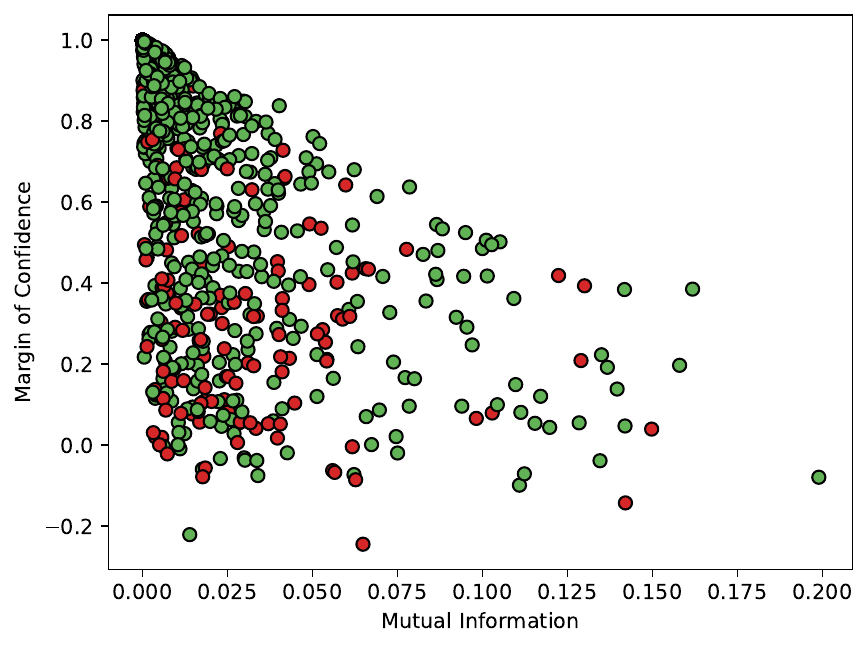}
    \caption{Bivariate scatter plot ($\mathbb{M}$ vs. $\mathbb{I}$) of correct (green) and incorrect (red) predictions for Cycle $8$.}
    \label{fig:mutual_moc_scatter}
\end{figure}

The $\mathbb{M}$ values were then plotted against the $\mathbb{I}$ values in Figure \ref{fig:mutual_moc_scatter} with point colors indicating whether the prediction was correct. There is a clear cluster of correct predictions in the northeast, corresponding to high $\mathbb{M}$ values with low $\mathbb{I}$ values. This is desirable, since a large number of \textit{correct} predictions are made with a relatively \textit{low} uncertainty. Moving down along the $\mathbb{M}$ axis, more incorrect predictions begin to populate. For the majority of these incorrect points, they exhibit low to moderate values of $\mathbb{I}$. There is a clear trend in incorrect predictions to increase in frequency with decreasing $\mathbb{M}$. This is another desirable occurrence, as \textit{incorrect} predictions would ideally display \textit{higher} levels of uncertainty. In general, the incorrect predictions do not seem well-correlated with $\mathbb{I}$, with a majority occurring below $\mathbb{I} = 0.05$. 

A fixed threshold on $\mathbb{M}$, termed $\widehat{\mathbb{M}}$, was then considered to compute uncertainty metrics from the segmented ``certain'' and ``uncertain'' predictions. Assessing if each point is certain or uncertain for a given threshold, tallies were taken for the number of correct-certain ($N_{\text{cc}}$), incorrect-uncertain ($N_{\text{iu}}$), correct-uncertain ($N_{\text{cu}}$), and incorrect-certain ($N_{\text{ic}}$) predictions. Higher relative numbers of $N_{\text{cc}}$ and $N_{\text{iu}}$ are considered desirable. Once these were taken, the \textbf{uncertainty accuracy} ($\text{UA}$) was computed with Equation (\ref{eq:uncertainty_accuracy}). This is a measure of the overall accuracy of uncertainty estimation proposed by Asgharnezhad et al. \cite{asgharnezhad2022objective} and is the simple ratio of desirable prediction outcomes compared to the total number of predictions. 

\begin{equation} \label{eq:uncertainty_accuracy}
    \text{UA}(\widehat{\mathbb{M}}) = \frac{N_{\text{cc}}+N_{\text{iu}}}{N_{\text{cc}}+N_{\text{iu}}+N_{\text{ic}}+N_{\text{cu}}}
\end{equation}

Likewise, sensitivity and specificity in the context of uncertainty can be computed using these tallies \cite{milanes2021monte}\cite{mobiny2021dropconnect}. Following the recommendations of \cite{asgharnezhad2022objective}, a choice of threshold can then be made by assessing trade-offs when the sensitivity and specificity are plotted over various thresholds. For this study, the threshold was set to $\widehat{\mathbb{M}} = 0.4$.

Similar to the evaluation of the base model's performance, confusion matrix-style metrics can be employed to evaluate the MCD-model's performance in an uncertainty frame. The ratio of correct-certain predictions ($R_{\text{cc}}$) and incorrect-uncertain predictions ($R_{\text{iu}}$) can now be computed with Equations (\ref{eq:rcc}) and (\ref{eq:riu}). The $R_{\text{cc}}$ value is the proportion of correct-certain predictions compared to all certain predictions made. The $R_{\text{iu}}$ value corresponds to the proportion of incorrect-uncertain predictions compared to all incorrect predictions made. These computed values along with their complements are provided in Table \ref{tab:uncertainty_metrics}.

\begin{equation} \label{eq:rcc}
    R_{\text{cc}} = \frac{N_{\text{cc}}}{N_{\text{cc}}+N_{\text{ic}}}
\end{equation}
\begin{equation} \label{eq:riu}
    R_{\text{iu}} = \frac{N_{\text{iu}}}{N_{\text{iu}}+N_{\text{ic}}}
\end{equation}

Assessing these uncertainty metric values, it is clear that a large majority of correct predictions made are certain. With an $R_{\text{cc}}$ value of $0.947$, just $6.4\%$ of correct predictions are classified as uncertain. This is again desirable, since an ideal model should be highly confident in the correct predictions made. When considering the ratio of incorrect-uncertain predictions, a value of $0.412$ suggests that while the model is confident in correct predictions, it  has a propensity to be confident in $36.5\%$ of incorrect predictions. This is clearly an undesirable quality in the model, but one that could potentially be mitigated with the use of a larger and more representative dataset or the implementation of calibration techniques such as Platt and/or temperature scaling \cite{guo2017calibration}.

\begin{table}[ht!]
	\centering
	\begin{tabular}{lc}
		\hline
		\textbf{Metrics}       & \textbf{Values} \\ \hline
		$\text{UA}$   & $0.873$  \\
		$R_{\text{cc}}$    & $0.947$ \\
        $R_{\text{ic}}$   & $0.365$ \\
		$R_{\text{cu}}$    & $0.053$ \\
        $R_{\text{iu}}$   & $0.635$ \\ \hline
	\end{tabular}	
	\caption{Uncertainty metrics using the $\mathbb{M}$ threshold of $0.4$.}
	\label{tab:uncertainty_metrics}
\end{table}

Overall, the implementation of MCD over the course of this section has allowed for extensive quantification of the model's uncertainties in multiple perspectives. The consistency and repeatability of the classifier's predictions were first considered, with every metric performing within $0.01$ of the base model. The reliability diagram exhibiting a Brier score of $0.052$ then suggested that the model is well-calibrated. On an assembly-wise level, an accuracy of $94.5\%$ was noted with a high degree of confidence in positive predictions at $254$ EFPD, when CIPS was most prevalent. 
Uncertainty measures, most notably the $\mathbb{M}$ and $\mathbb{I}$ values between true label and model parameters, were computed to display expected trends of correct predictions increasing in frequency with increasing $\mathbb{M}$ and decreasing $\mathbb{I}$. Finally, the $\mathbb{M}$ values were thresholded to provide uncertainty metrics; $94.7\%$ of correct predictions were found to be certain while $63.5\%$ of incorrect predictions were uncertain. Overall, these results indicate that this approach could potentially be useful in the prediction of CIPS instances on an assembly-wise level.

\section{Discussion} \label{sec:discussion}

The above results demonstrate that a surrogate model can effectively predict instances of CIPS in CNS-1 by using a combination of core model and measurement data from previous cycles. When considering the deployment of this approach to an operational capacity in the frame of industry, practical concerns must be addressed. The most important is defining the scope of its use case, which also is relevant from a regulatory standpoint.

It is unlikely that such a framework would be of use as a means of evaluating a finalized core design, since utilities and vendors have established methods that are well-accepted in the regulatory environment. This method's use, however, is to instead be a first-order estimation tool to better-inform core designers while they are in the phase of prototyping loading patterns. This process often requires several potential designs to be run, and with the current physics-based crud tools, can take a significant amount of time per design as noted in Section \ref{subsec:timing}. Rather than each prospective design being assessed with these conventional tools, the approach described in this study would effectively eliminate that time cost by providing accurate real-time estimates even if they are lower-fidelity. By being involved in optimization step, with a finalized design still being evaluated by conventional means, it would not be involved in any final safety-related workflow processes or formal analyses, a benefit from a regulatory standpoint.

From a practical standpoint, a well-trained model could be implemented into a nodal core simulator \cite{che2022machine}, many of which have a graphical user interface to place assemblies around a quarter or eighth-core to obtain a heat map of information like power peaking factors. With the CIPS model's inferences made on the scale of milliseconds, it would be fast enough to provide CIPS potential on an assembly-wise basis as the assemblies are being configured. A potential visualization of this would be a heat map overlay such as the visualization in Figure \ref{fig:quarter_254efpd} of Section \ref{subsubsec:assemblywise_confidence}. Not only would this provide the predicted probability of each assembly's CIPS state, but it would also provide the predictions' confidences to help better-inform the core designer.

In terms of progressing a model built on this methodology to a mature state, substantive validation is required not only at CNS-1 but to also determine if it generalizes to other reactors. This specific model's ability to do this is unknown, as each reactor's CIPS behavior is typically unique. The model in this study is specifically calibrated to CNS-1 to support the objective to construct and validate the methodology itself. Improving performance and determining its generalization would require more data from CIPS-affected cycles in different reactors. The data problem is most challenging aspect of implementing this approach, as both operating data and core design data are difficult to obtain. Accomplishing this would likely require partnership with a utility and/or reactor vendor with access to large amounts of this data. Once data is obtained, however, the model can be rapidly updated to improve its performance and reliability on specific reactor units, as well as in exploring the transferability of shared features between different units with techniques such as transfer learning.

\section{Conclusions}\label{sec:conclusion}

In this study, a methodology was proposed for using 3D convolutional neural networks to predict instances of crud-induced power shift (CIPS) in pressurized water reactors. Using data from Cycles 7 and 8 of the Catawba Nuclear Station Unit 1, the network was designed to make predictions using boron concentration, core exposure, and 3D pin power maps. This methodology used a classification approach to predict whether an assembly location at a given time would be CIPS-positive or CIPS-negative. 

The training time for the classifier was on average $215.9$ minutes using a set of four Apple M1 cores, with an inference time of $17.2$ milliseconds to yield predictions for the two cycles post-training. The overall accuracy of the concatenated test predictions was $92.3\%$ with an $F_1$ score and Matthews Correlation Coefficient of $0.798$ and $0.752$ respectively. These values indicate a high positive correlation between predictions and true class labels. The percentage of assemblies affected by CIPS at a given time was then computed for predictions and true labels to produce a set of curves over Cycle 8. The classifier was shown to effectively predict when the onset of CIPS occurs, when the peak of CIPS occurs along with its magnitude, and captures the overall trajectory of CIPS development over Cycle 8.

Uncertainty quantification of the classifier was then performed by implementing Monte Carlo dropout (MCD), an easy-to-use technique that produces output distributions for the input entries. The consistency and repeatability of the classifier's predictions were first considered, with every metric performing within $0.01$ of their non-MCD counterparts. Both the base model and MCD equivalent's calibrations were compared by using reliability diagrams, where the MCD approach was observed with slightly improved calibration compared to the base. On an assembly-wise level, an accuracy of $94.5\%$ was noted with a high degree of confidence in positive predictions at $254$ EFPD, when CIPS was most prevalent. Uncertainty measures, most notably the margin of confidence and mutual information between true label and model parameters, were computed to display expected trends of correct predictions increasing in frequency with increasing margin of confidence and decreasing mutual information. Finally, the margin of confidence was thresholded to provide uncertainty metrics; $94.7\%$ of correct predictions were found to be certain while $63.5\%$ of incorrect predictions were uncertain. 

Future work was also discussed, with particular emphasis on the steps necessary for such a method to be deployed to assist in core design in both CNS-1 and new reactors. With the results enclosed in this study, this approach has the potential to be useful in a core design setting for the rapid prototyping of loading patterns by providing information about potential CIPS effects.

\section*{Acknowledgment}

This work was supported by Duke Energy through the Consortium for Nuclear Power (CNP) at the Department of Nuclear Engineering of North Carolina State University.

\bibliography{./bibliography.bib}

\begin{thebibliography}{10}
\bibitem{kendall2017uncertainties}
A.~Kendall, Y.~Gal, What uncertainties do we need in bayesian deep learning for computer vision?, Advances in Neural Information Processing Systems 30.
\end{thebibliography}

\end{document}